\documentclass[useAMS]{mn2e}
\usepackage{graphicx}

\def\lesssim{\,\lower2truept\hbox{${<\atop\hbox{\raise4truept\hbox{$\sim$}}}$}\,}
\def\gtrsim{\,\lower2truept\hbox{${>\atop\hbox{\raise4truept\hbox{$\sim$}}}$}\,}
    
    \def\smallskip{\vskip 6pt}
    
    \def\M12{${\rm M_{12}}$}
    
\newcommand{\lsim}{\mathrel{\rlap{\raise -.3ex\hbox{${\scriptstyle\sim}$}}%
                   \raise .6ex\hbox{${\scriptstyle <}$}}}%
\newcommand{\gsim}{\mathrel{\rlap{\raise -.3ex\hbox{${\scriptstyle\sim}$}}%
                   \raise .6ex\hbox{${\scriptstyle >}$}}}%

\title[Evolution of spheroidal galaxies]{Observational tests of the evolution of spheroidal galaxies}
\author[Silva et al.]{L. Silva$^{1}$, G. De Zotti$^{2,4}$, G. L. Granato$^{2,4}$,
R. Maiolino$^{3}$ \& L. Danese$^{4}$ \\
$^{1}$INAF - Osservatorio astronomico di Trieste, Via Tiepolo 11, I-34131 Trieste, Italy\\
$^{2}$INAF - Osservatorio astronomico di Padova, Vicolo
dell'Osservatorio 5, I-35122 Padova,
Italy \\
$^{3}$INAF - Osservatorio astrofisico di Arcetri, Largo E. Fermi 5, I-50125 Firenze, Italy\\
$^{4}$SISSA, Via Beirut 2, I-34014 Trieste, Italy}

\begin{document}
\date{Accepted. Received}
\pagerange{\pageref{firstpage}--\pageref{lastpage}} \pubyear{2004}
\maketitle

\label{firstpage}

\begin{abstract}
Granato et al. (2004a) have elaborated a physically grounded model
exploiting the mutual feedback between star-forming spheroidal
galaxies and the active nuclei growing in their cores to overcome,
in the framework of the hierarchical clustering scenario for
galaxy formation, one of the main challenges facing such scenario,
i.e. the fact that massive spheroidal galaxies appear to have
formed much earlier and faster than predicted by previous
semi-analytical models, while the formation process was slower for
less massive objects. After having assessed the values of the two
parameters that control the effect of the complex and poorly
understood radiative transfer processes on the time-dependent
spectral energy distributions (SEDs), we have compared the model
predictions with a variety of IR to mm data. Our results support a
rather strict continuity between objects where stars formed
(detected by (sub)-mm surveys) and evolved massive early-type
galaxies, indicating that large spheroidal galaxies formed most of
their stars when they were already assembled as single objects.
The model is remarkably successful in reproducing the observed
redshift distribution of $K\le 20$ galaxies at $z>1$, in contrast
with both the classical monolithic models (which overestimate the
density at high-$z$) and the semi-analytic models (that are
systematically low), as well as the ratio of star-forming to
passively evolving spheroids and the counts and redshift
distributions of Extremely Red Objects (EROs), although the need
of a more detailed modelling of the star formation history and of
dust geometry is indicated by the data. The model also favourably
compares with the ISOCAM $6.7\,\mu$m counts, with the
corresponding redshift distribution, and with Spitzer/IRAC counts,
which probe primarily the passive evolution phase, and with the
(sub)-mm SCUBA and MAMBO data, probing the active star-formation
phase. The observed fraction of $24\,\mu$m selected sources with
no detectable emission in either the $8\,\mu$m or $R$-band (Yan et
al. 2004a) nicely corresponds to the predicted surface density of
star-forming spheroids with $8\,\mu$m fluxes below the detection
limit. Finally, distinctive predictions for the redshift
distributions of $24\,\mu$m sources detected by Spitzer/MIPS
surveys are pointed out.
\end{abstract}


\begin{figure*}
\centering
\includegraphics[width=16 truecm]{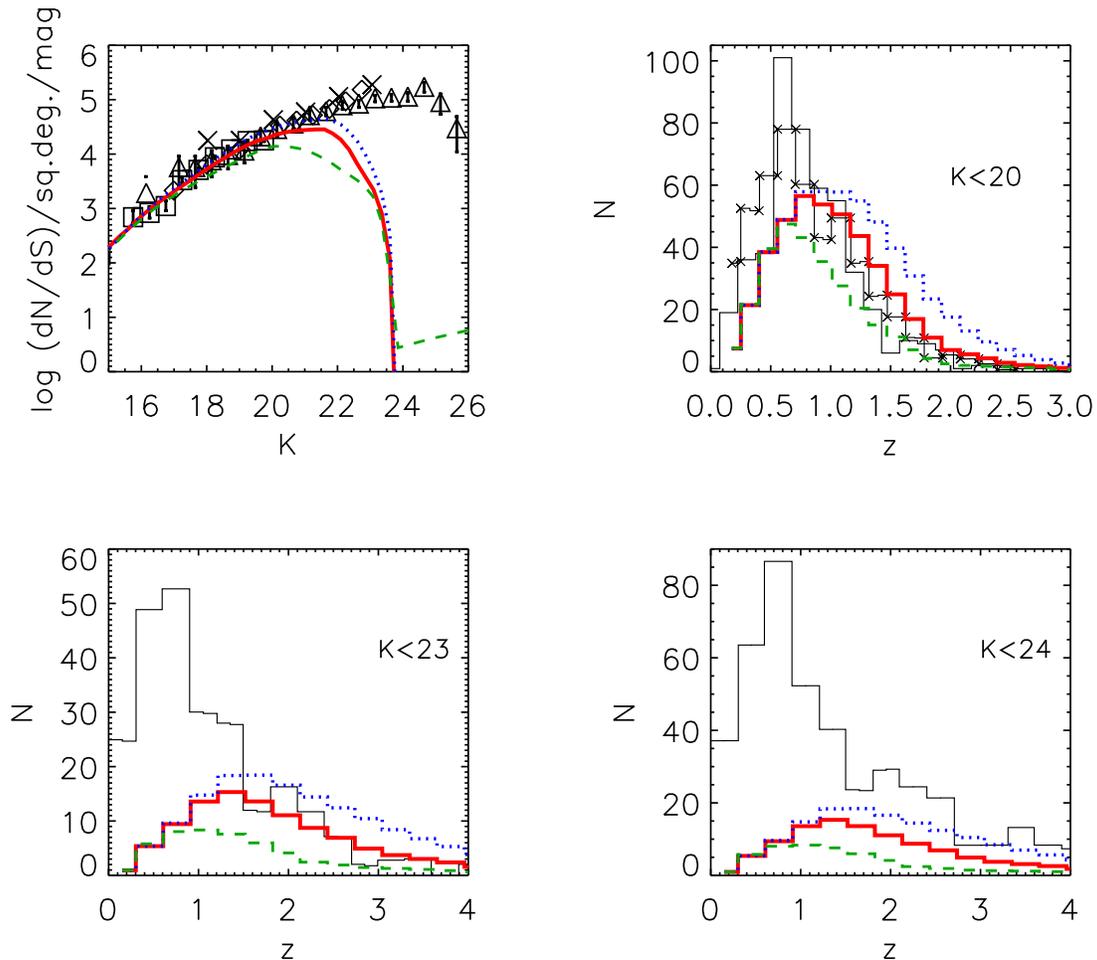}
\caption{Contributions of spheroidal galaxies to the $K$-band
number counts and redshift distributions of galaxies brighter than
$K=20$, 23, 24. The dashed, solid and dotted lines refer to $k=3$,
30, $300\, M_\odot^{1/3}$ respectively (see Sect. \ref{sect:SED}).
We adopt $k=30 M_\odot^{1/3}$ as our reference value.  Data for K
band counts are from Moustakas et al. (1997), Kochanek et al.\
(2001), Saracco et al.\ (2001), Totani et al.\ (2001), Cimatti et
al.\ (2002). In the upper-right panel (the K$<20$ redshift
distribution), the thin solid and the thin solid with crosses
histograms (the redshift bin $\Delta z$ is 0.15) are respectively
from Cimatti et al.\ (2002) and Somerville et al.\ (2004) (scaled
to the same area of the K20 survey). In the lower panels the thin
continuous histograms ($\Delta z$=0.3) are from Kashikawa et al.
(2003).} \label{cK}
\end{figure*}

\section{Introduction}

The standard Lambda Cold Dark Matter ($\Lambda$CDM) cosmology is a
well established framework to understand the hierarchical assembly
of dark matter (DM) halos. Indeed, it has been remarkably
successful in matching the observed large-scale structure. However
the complex evolution of the baryonic matter within the potential
wells determined by DM halos is still an open issue, both on
theoretical and on observational grounds.

Full simulations of galaxy formation in a cosmological setting are
far beyond present day computational possibilities. Thus, it is
necessary to introduce at some level rough parametric
prescriptions to deal with the physics of baryons, based on
sometimes debatable assumptions (e.g.\ Binney 2004). A class of
such models, known as semi-analytic models, has been extensively
compared with the available information on galaxy populations at
various redshifts (e.g.\ Lacey et al.\ 1993; Kauffmann, White \&
Guiderdoni, 1993; Cole et al.\ 1994; Kauffmann et al.\ 1999;
Somerville \& Primack 1999; Cole et al.\ 2000; Granato et al.\
2000; Benson et al.\ 2003; Baugh et al. 2004).

The general strategy consists in using a subset of observations to
calibrate the many model parameters providing a heuristic
description of baryonic processes we don't properly understand.
Besides encouraging successes, current semi-analytic models have
met critical inconsistencies which seems to be deeply linked to
the standard recipes and assumptions. These problems are in
general related to the properties of elliptical galaxies, such as
the color-magnitude and the [$\alpha$/Fe]-M relations (Cole et al.
2000; Thomas 1999; Thomas et al. 2002), and the statistics of
sub-mm and deep IR selected (I- and K-band) samples (Silva 1999;
Chapman et al. 2003; Kaviani et al. 2003; Daddi et al. 2004;
Kashikawa et al. 2003; Poli et al. 2003; Pozzetti et al. 2003;
Somerville et al. 2004). Discrepancies between the observationally
estimated merger rate evolution and that predicted by
semi-analytic models have also been reported (Conselice et al.
2003).

However, the general agreement of a broad variety of observational
data with the hierarchical scenario and the fact that the observed
number of luminous high-redshift galaxies, while substantially
higher than predicted by standard semi-analytic models, is
nevertheless consistent with the number of sufficiently massive
dark matter halos, indicates that we may not need alternative
scenarios, but just some new ingredients or assumptions (see e.g.\
Baugh et al.\ 2004; Tecza et al. 2004).

Previous work by our group (Granato et al.\ 2001; Romano et al.\
2002; Granato et al.\ 2004a) suggests that a crucial ingredient is
the mutual feedback between spheroidal galaxies and active nuclei
at their centers. Granato et al.\ (2004a, henceforth GDS04)
presented a detailed physically motivated model for the early
co-evolution of the two components, in the framework of the
$\Lambda$CDM cosmology.

In this paper, we present a comparison of the model with the
observed number counts and redshift distributions in several
near-IR (NIR) to (sub)-mm bands.

In Section~ \ref{sec:mod} we give a short overview of the GDS04
model for spheroidal galaxies, and we discuss the determination of
the parameters controlling the time-dependent spectral energy
distributions (SEDs) of spheroidal galaxies. In Section~
\ref{sect:comdat} , the model predictions are compared with a
variety of recent data. The main conclusions are summarized and
discussed in Section~\ref{sect:conclu}.

We adopt the ``concordance'' cosmological model, with
$\Omega_m=0.3$, $\Omega_\Lambda=0.7$,
$H_0=70\,\hbox{km}\,\hbox{s}^{-1}\,\hbox{Mpc}^{-1}$.

\section{Model description and assessment of the parameters}
\label{sec:mod}

\subsection{The GDS04 model}
\label{sec:gds04}

While referring to GDS04 for a full account of the model
assumptions and their physical justification, we provide here, for
the reader's convenience, a brief summary of its main features.

The model follows with simple, physically grounded recipes and a
semi-analytic technique the evolution of the baryonic component of
proto-spheroidal galaxies within massive dark matter (DM) halos
forming at the rate predicted by the standard hierarchical
clustering scenario within a $\Lambda$CDM cosmology. The main
novelty with respect to other semi-analytic models is the central
role attributed to the mutual feedback between star formation and
growth of a super massive black hole (SMBH) in the galaxy center.
Further relevant differences are the assumption that the large
scale angular momentum does not effectively slow down the collapse
of the gas and the star formation in massive halos virialized at
high redshift, and the allowance for a clumping factor, $C\simeq
20$, speeding up the radiative cooling so that even in very
massive halos ($M_{\rm vir}\sim 10^{13}\,M_\odot$), the gas,
heated to the virial temperature, can cool on a relatively short
($\sim 0.5$--$1\,$Gyr) timescale, at least in the dense central
regions.

The GDS04 prescriptions are assumed to apply to DM halos
virializing at $z_{\rm vir} \gsim 1.5$ and $M_{\rm vir} \gsim 4
\times 10^{11} M_\odot$. These cuts are meant to crudely single
out galactic halos associated with spheroidal galaxies. Disk (and
irregular) galaxies are envisaged as associated primarily to halos
virializing at $z_{\rm vir} \lsim 1.5$, some of which have
incorporated most halos less massive than $4 \times
10^{11}\,M_\odot$ virializing at earlier times, that may become
the bulges of late type galaxies. However, the model does not
address the formation of disk (and irregular) galaxies and these
objects will not be considered in this paper. On the other hand,
we do not expect that proto-spheroidal galaxies in the
star-forming phase have already a spheroidal shape; rather they
are likely to have an irregular, complex, disturbed morphology.

The kinetic energy fed by supernovae is increasingly effective,
with decreasing halo mass, in slowing down (and eventually
halting) both the star formation and the gas accretion onto the
central black hole. On the contrary, star formation and black hole
growth proceed very effectively in the more massive halos, giving
rise to the bright SCUBA phase, until the energy injected by the
active nucleus in the surrounding interstellar gas unbinds it,
thus halting both the star formation and the black hole growth
(and establishing the observed relationship between black hole
mass and stellar velocity dispersion or halo mass). Not only the
black hole growth is faster in more massive halos, but also the
feedback of the active nucleus on the interstellar medium is
stronger, to the effect of sweeping out such medium earlier, thus
causing a shorter duration of the active star-formation phase. As
a result, in keeping with the previous proposition by Granato et
al.\ (2001), the physical processes acting on baryons reverse the
order of formation of spheroidal galaxies with respect to the
hierarchical assembling of DM halos, leading to the {\it
Anti-hierarchical Baryon Collapse} (ABC) scenario.

\begin{figure}
\centering
\includegraphics[width=9 truecm]{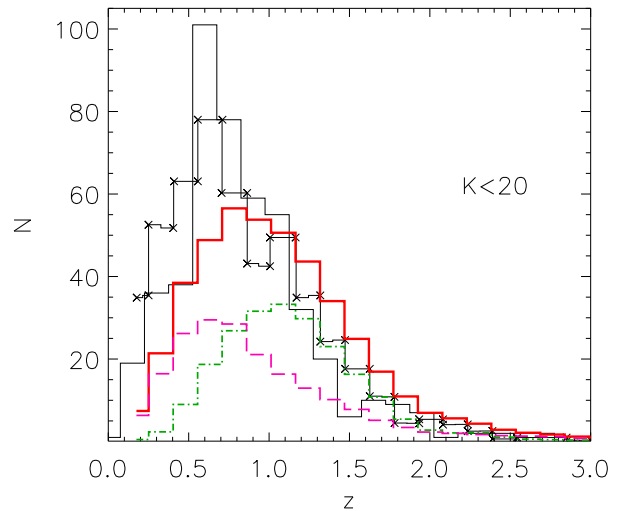}
\caption{Contribution of star forming (dot-dashed line) and
passively evolving (long-dashed line) spheroids to the $K<20$
redshift distribution (the same as in Fig.\ \ref{cK}). The thick
solid line is the model total, the thin line the Cimatti et al.\
(2002) data, and the thin line with crosses the data from
Somerville et al.\ (2004) scaled to the same area. Note that
Somerville et al. give magnitudes in the AB system
($K_{AB}=K+1.85$). } \label{cK2}
\end{figure}

\begin{figure*}
\centering
\includegraphics[width=16 truecm]{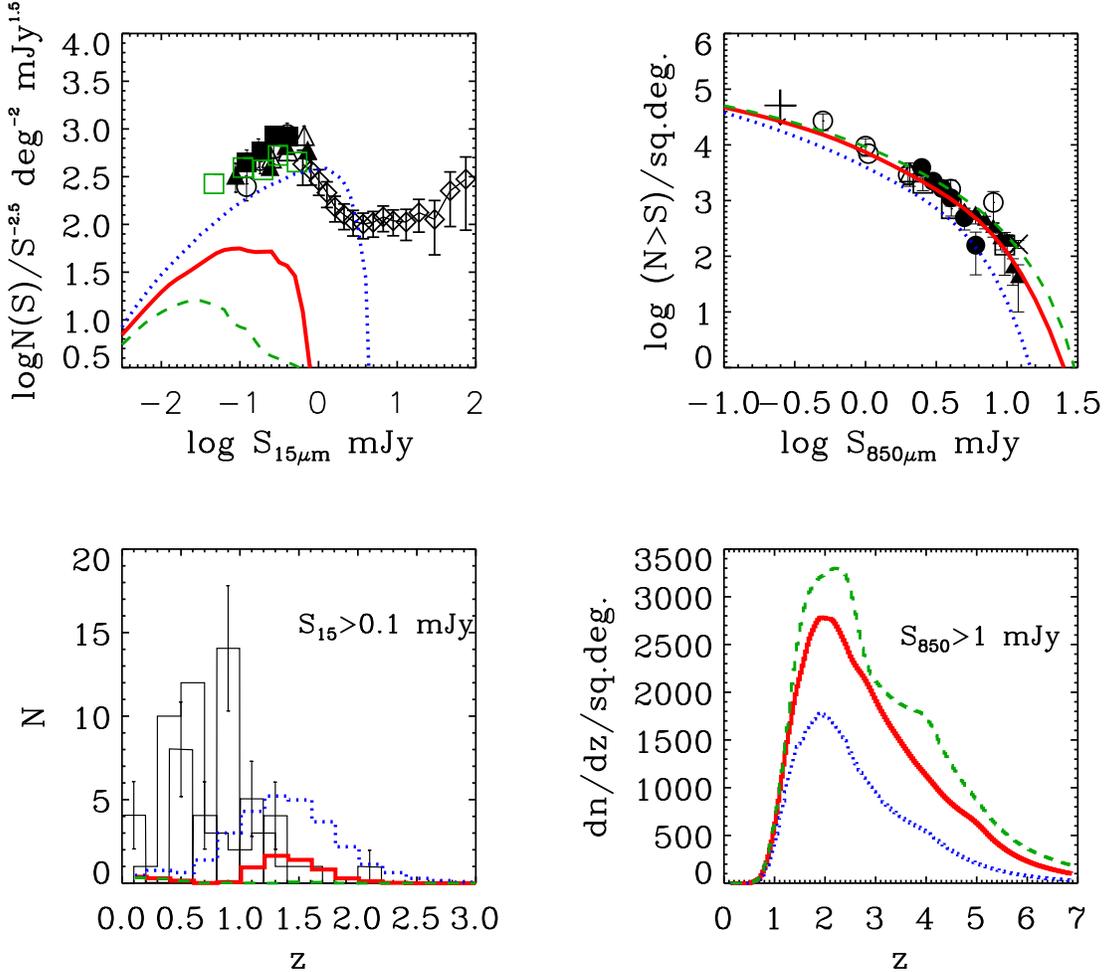}
\caption{Possible contributions of spheroids to the observed 15
and 850 $\mu$m number counts and redshift distributions ($S_{15
\mu{\rm m}}>0.1$mJy within $6\times 10^{-3}$ deg$^2$ and $\Delta
z=0.2$; $S_{850 \mu m}>1\,$mJy). The dotted, solid and dashed
lines refer to $\tau_{\rm MC}(1 \mu m) \simeq 10$, 60, 120
respectively. The $15\,\mu$m data are from Elbaz et al.\ (1999),
Gruppioni et al.\ (2002),  Elbaz et al.\ (2002, thin histogram
with error bars), Franceschini et al.\ (2003, thin histogram),
Metcalfe et al.\ (2003). The $850\,\mu$m data are from Hughes et
al.\ (1998), Barger, Cowie, \& Sanders (1999), Blain et al.\
(1999), Eales et al.\ (2000), Borys et al.\ (2002), Chapman et
al.\ (2002). A low $\tau_{\rm MC}(1 \mu m)$ implies an excessive
contribution from spheroids to the $15\,\mu$m counts, which are
dominated by late-type, mainly starburst, galaxies at $z \lesssim
1$ (Hammer et al. 2004). On the other hand, a very high $\tau_{\rm
MC}(1 \mu m)$ overpredicts the $850 \mu$m counts. Our reference
model has $\tau_{\rm MC}(1 \mu{\rm m})=60$. See text for details.}
\label{c15_850}
\end{figure*}

\subsection{Spectral energy distribution of spheroidal galaxies }
\label{sect:SED}

As in GDS04, we compute the SEDs of spheroidal galaxies using our
code GRASIL, described in Silva et al. (1998)\footnote{The code is
available at {\it http://adlibitum.oat.ts.astro.it/silva/
default.html} or at {\it http://web.pd.astro.it/granato}}. GRASIL
computes the time-dependent UV to radio SED of galaxies, given
their star formation and chemical enrichment histories (derived as
described in GDS04), and with a state of the art treatment of dust
reprocessing. The latter point is fundamental, since during the
phase of intense star-formation, young stars are mixed with a huge
amount of gas, quickly chemically enriched and dust polluted. The
use of local templates (such as M82 or Arp220) may be
inappropriate in these extreme conditions.

One of the most important distinctive features of GRASIL is that
it has included, for the first time, the effect of {\it
differential extinction} of stellar populations (younger stellar
generations are more affected by dust obscuration), due to the
fact that stars form in a denser than average environment, the
molecular clouds (MCs), and progressively get rid of them.

Radiative transfer results depend on the poorly known geometry of
the system. Broadly speaking, predictions of fluxes during the
active star forming phase become more and more affected by the
ensuing model uncertainties at shorter and shorter wavelengths. In
GDS04 we presented predictions for the dust-enshrouded SCUBA
galaxies, or for the essentially dust-free passively evolving
galaxies; in both cases, the results were insensitive to the
details of the dust distribution. On the other hand, key
information on the early evolution of spheroidal galaxies is
provided by near-IR (particularly $K$-band) and mid-IR (ISO and
Spitzer) data. To test the model against them we need a more
careful investigation of the dependence of our results on the
structure of galaxies.

In general, the GRASIL SEDs for spheroidal systems depend on five
structural parameters: the fraction $f_{\rm MC}$ of gas in the
form of MCs rather than in the diffuse ISM (cirrus), the optical
depth of MCs for a source at their center $\tau_{\rm MC}$
(conventionally given at 1 $\mu$m hereafter), the escape timescale
of newly born stars from MCs, $t_e$, and the core radii of the
distributions of diffuse dust, $r_c^c$, and of stars outside MCs,
$r_c^*$, assumed to follow a King (1972) law. In the following we
will take $r_c^c=r_c^*$, and we will drop the superscript.

However, in the conditions envisaged here and for the purposes of
the present work, the results are significantly affected only by
$\tau_{\rm MC}$. This is proportional to the dust to gas mass
ratio $\delta$ and to $M_{\rm MC}/R_{\rm MC}^2$, $M_{\rm MC}$ and
$R_{\rm MC}$ being the typical mass and radius of molecular
clouds, respectively. As for $\delta$, we have simply assumed that
it increases linearly with the gas metallicity (e.g., Dwek 1998)
and is $\simeq 1/100$ (the Galactic value) for solar metallicity.
Since the chemical evolution of the galaxy is computed
self-consistently, the true parameter is $M_{\rm MC}/R_{\rm
MC}^2$, but we prefer to make reference to the value of $\tau_{\rm
MC}$ for solar metallicity (and thus $\delta \simeq 1/100$). We
have checked that all the results presented here, except those on
Extremely Red Objects (EROs, see Sect.~\ref{sec:ero}), are very
little affected by variations, within rather wide ranges, of the
other GRASIL parameters.

\begin{figure}
\centering
\includegraphics[width=8.5truecm]{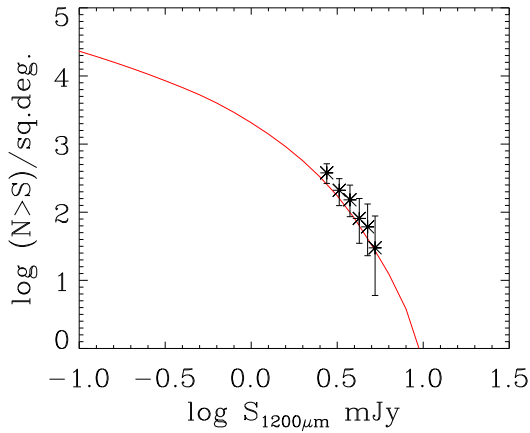}
\caption{Integral counts of star-forming spheroidal galaxies at
$1200 \mu$m predicted by our reference model, compared with MAMBO
data from Greve et al.\ (2004).} \label{c1200}
\end{figure}

\begin{figure}
\centering
\includegraphics[width=9truecm]{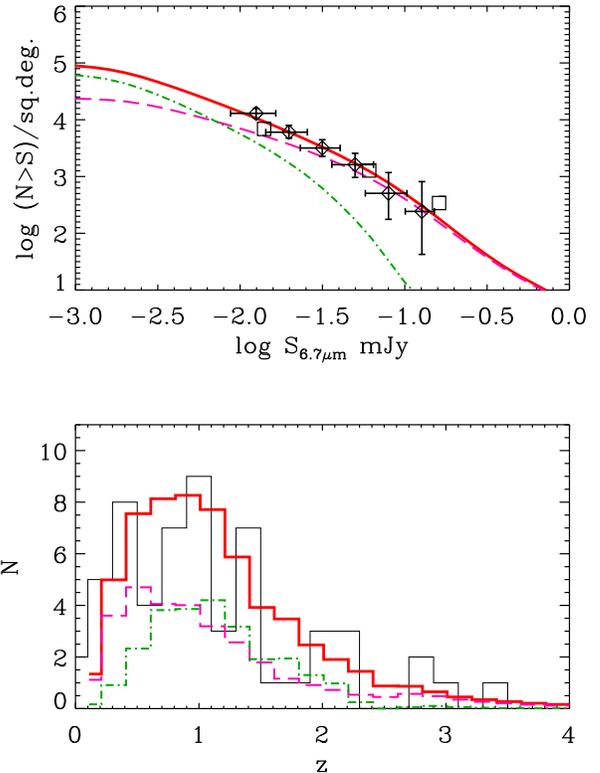}
\caption{$6.7 \mu$m counts and redshift distribution of sources
with $S_{6.7}>10\mu$Jy within $16$ arcmin$^2$. The thick solid
line shows the prediction of our reference model (i.e. $k=30$
$M_\odot^{1/3}$). The contributions of active and passive
spheroids are represented by the dot-dashed and long-dashed lines
respectively. The thin histogram in the lower panel shows the
redshift distribution (with $\Delta z=0.2$) observationally
estimated by Sato et al. (2004). The observed counts are from
Metcalfe et al.\ (2003, squares) and Sato et al. (2003,
diamonds).} \label{c6.7std}
\end{figure}

\begin{figure*}
\centering
\includegraphics[width=16 truecm]{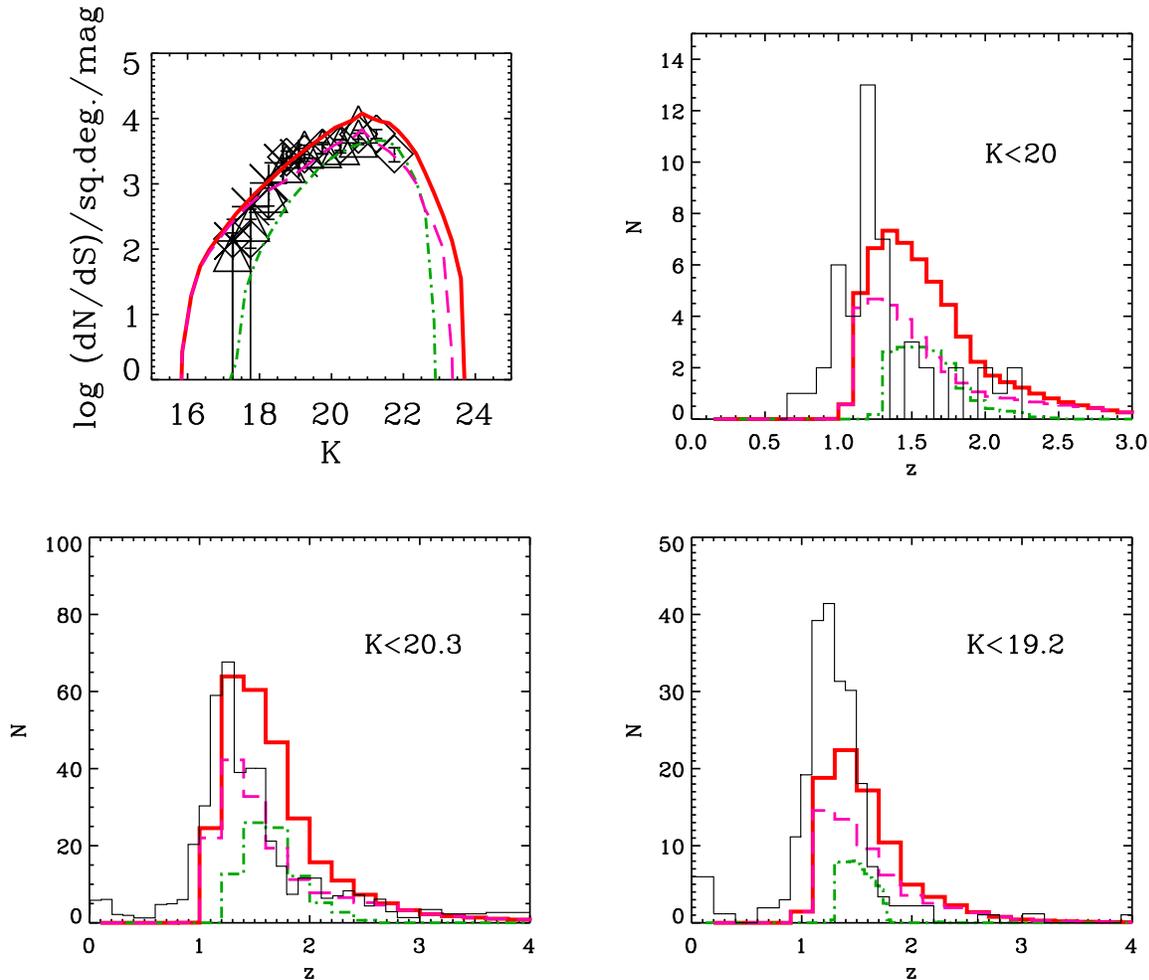}
\caption{$K$-band counts and redshift distributions for EROs
($R-K>5$). The thick solid line is the total predicted by our
model, while the contributions of active and passive objects are
shown by dot-dashed and long dashed lines respectively. Data from
Daddi et al.\ (2000), Miyazaki et al.\ (2003), Roche et al.\
(2002, 2003), Cimatti et al.\ (2003). The redshift histograms are
for $\Delta z=0.1$ for $K <20$, and for $\Delta z=0.2$ in the two
lower panels.} \label{counts_ero}
\end{figure*}

However, an additional parameter is required to circumvent the
problem that, for stars already outside their parent molecular
clouds, GRASIL takes into account only the relatively small
optical depth due to cirrus, neglecting that due to other MCs
present along the line of sight. This is not necessarily accurate
in the extreme conditions occurring during the fast star forming
phase of spheroids, particularly in the dense central regions. In
general, the mean optical depth to a star outside its parent MC is
$\tau_{\rm outside}=\tau_{\rm cirrus}+ (4/3) \bar{N} \tau_{\rm
MC}$, where we have assumed a uniform dust distribution inside
MCs, and $\bar{N}$ is mean number of MCs along its line of sight.
\begin{equation}
\bar{N} = S_f {M_{\rm gas, MC}\over M_{\rm MC}}\frac{R_{\rm
MC}^2}{r_c^2} \propto {M_{\rm gas, MC}}{M_{\rm vir}^{-2/3}}\ ,
\label{barN}
\end{equation}
where $S_f$ is a shape factor and we have assumed $r_c \propto
M_{\rm vir}^{1/3}$, $M_{\rm vir}$ being the virial mass of the
galaxy. Whenever $\bar{N}<<1$, the optical depth drops from
$\tau_{\rm MC}$ to $\tau_{\rm cirrus}$ ($\ll 1$ at IR wavelengths)
when the stars move out of their parent MCs. This condition is
usually met in nearby galaxies. On the contrary, in forming
spheroids $\tau_{\rm outside}$ can be substantially larger than
$\tau_{\rm cirrus}$. The shape factor $S_f$ depends on the
relative distribution of stars and MCs, and its value is also a
function of the age of the stellar population under consideration:
younger stars (but old enough to be out of molecular clouds) are
on average closer to regions rich in MCs.

Thus, an accurate calculation of $\bar{N}$ is extremely difficult,
and we are forced to parameterize our ignorance. Also, the present
release of GRASIL does not allow us to deal with the radiation
transfer through the distribution of MCs. On the other hand, such
detailed treatment is not crucial for our present purposes since
whenever $\tau_{\rm outside}\ge \tau_{\rm MC}$ essentially all the
starlight is reprocessed by dust. We have crudely approximated
$\tau_{\rm outside}$ as a step function: $\tau_{\rm
outside}=\tau_{\rm cirrus}$ for $M_{\rm gas, MC} \le k M_{\rm
vir}^{2/3}$ and $\tau_{\rm outside}=\tau_{\rm cirrus}+ \tau_{\rm
MC}$ otherwise, treating $k$ as an adjustable parameter. In order
to take into account the dust heating by the absorbed starlight,
in practice we have kept the stars within their MCs for the full
duration of the starburst, for galaxies where $M_{\rm gas, MC} > k
M_{\rm vir}^{2/3}$.

The parameter $k$ affects mostly the results at $\lambda \lesssim
10 \mu$m. The $K$-band counts and redshift distributions are best
reproduced setting $k \sim 30 M_\odot^{1/3}$ (Fig.~\ref{cK}). Note
that spheroids dominate the redshift distributions at substantial
redshifts, while leaving room for other populations at low
redshifts. The figure shows also the effect of varying $k$. In
particular, a low value of $k$ ($k \simeq 10$, not shown), would
produce an almost perfect match of the $K<20$ redshift
distribution for $z \gtrsim 1$, but the contributions at $K<23$
and $K<24$ would decrease. A more detailed treatment of the
complex behavior of the optical depth to stars outside MCs could
improve the overall match, and will be considered for future
releases of GRASIL.

The results in Fig.~\ref{cK} refer to $t_e=0.01\,$Gyr, $f_{\rm
MC}=0.5$ and $r_c=0.1 (M_{\rm vir}/10^{12} M_\odot)^{1/3}$ kpc,
but change only marginally if $t_e$ and $f_{\rm MC}$ vary within
rather broad intervals ($5 \, {\rm Myr} \lesssim t_e\lesssim 0.5$
Gyr, $0.2 \lesssim f_{\rm MC} \lesssim 0.9$), and $r_c$ is
increased or decreased by a factor of a few.

As for $\tau_{\rm MC}$, Silva et al.\ (1998) found that the SEDs
of starburst galaxies are well reproduced by GRASIL with
$\tau_{\rm MC} \sim \mbox{a few tens}$ (a figure in reasonable
agreement with direct observations of Giant Molecular Clouds in
the Galaxy, for which $M_{\rm MC} \sim 10^6 M_\odot$, $r_{\rm MC}
\sim 15\,$pc). However, the physical conditions in high-$z$ star
forming spheroids may be very different from those encountered in
local systems, including starbursts and Ultra-Luminous IR Galaxies
(ULIRGs). Thus, we have explored the generous range $3 < \tau_{\rm
MC} < 200$. This parameter controls the contribution (if any) of
star-forming spheroidal galaxies to the bump in the $15\,\mu$m
ISOCAM counts below a few mJy. Although the observed counts are
affected by a considerable uncertainty, as shown by the spread of
results from different surveys, and the modelling is complicated
by the presence of PAH bands, they have been interpreted by
phenomenological models in terms of evolving starbursts, with a
minor contribution from spirals, particularly at the brighter flux
density levels (e.g. Franceschini et al.\ 2001; Gruppioni et al.\
2002; Lagache et al. 2003; Hammer et al. 2004). Thus, the
contribution from star-forming spheroids must be substantially
suppressed, implying $\tau_{\rm MC}\gsim 40$ at $1 \mu$m (see
Fig.~\ref{c15_850}). A low value of $\tau_{\rm MC}$, and,
correspondingly, a large contribution from spheroids to the
sub-mJy bump of the $15\,\mu$m counts (upper-left panel), would
entail a high-$z$ tail of the redshift distribution of sources
with $S_{15} > 0.1\,$mJy substantially larger than indicated by
current data (lower left-hand panel).

On the other hand, as illustrated by the right-hand panels of
Fig.~\ref{c15_850}, an exceedingly large value of $\tau_{\rm MC}$
would overproduce the $850\,\mu$m counts, while a very small value
would underproduce them. The ensuing lower limit to $\tau_{\rm
MC}$ is however model dependent, since the effect of decreasing it
may be compensated by a smaller value of the dust emissivity
index, that we take equal to 2.

In the following we set $\tau_{\rm MC}\simeq 60$ at $1 \mu$m, a
value yielding $850\,\mu$m counts and redshift distributions
consistent with the data, and a small, but non-negligible,
contribution to the sub-mJy $15\,\mu$m counts.

\section{Comparison with the data}
\label{sect:comdat}

\subsection{Counts and redshift distributions}
\label{sect:coured}

As discussed by Granato et al. (2001, 2004a), in the present
framework the SCUBA-selected galaxies are interpreted as being
mostly star-forming massive proto-spheroidal galaxies ($M_{\rm
vir} \geq 10^{11.6}\, M_\odot$). The values of the GRASIL
parameters derived above nicely fit the $850\,\mu$m SCUBA counts
(see Fig.~\ref{c15_850}, right-hand panels), and the latest MAMBO
$1200 \mu$m data (Fig.~\ref{c1200}). The model $850\,\mu$m counts
are almost undistinguishable from those already reported in
Fig.~12 of GDS04, corresponding to a somewhat different choice of
the parameters, and the redshift distributions are also
essentially unchanged. Indeed, the predictions in the sub-mm
region are not very sensitive to the precise values of GRASIL
parameters, as far as proto-spheroidal galaxies are optically
thick in the optical-UV. The predicted redshift distribution of
sources brighter than $1\,$mJy at $850\,\mu$m has a broad peak at
$z\sim 2$ with an extended tail towards higher redshifts, due to
star-forming spheroidal galaxies. As already remarked in GDS04,
this redshift distribution is in good agreement with the limited
information available so far. In particular, the median redshift
of the Chapman et al.\ (2003) sample ($S_{850\mu{\rm m}} \ge
5\,$mJy) is $2.4$, with a quartile range of $z=1.9-2.8$. The
median and quartile range for our model are $2.3$ and $1.8-3.0$.

\begin{figure}
\centering
\includegraphics[width=8.5truecm]{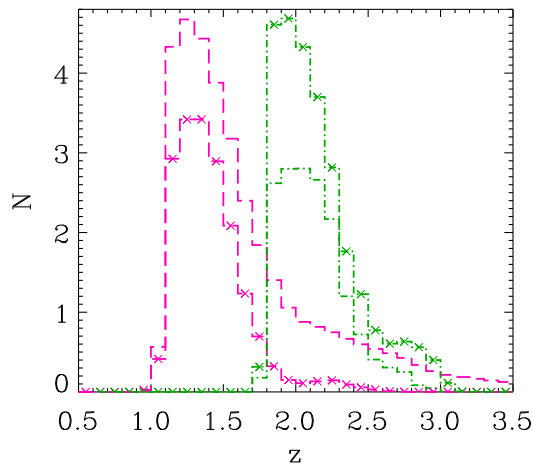}
\caption{Dependence of predicted redshift distribution of EROs in
the K20 survey on details of the model (see Fig. \ref{counts_ero},
upper right panel). The dashed line with crosses is obtained
adding to the reference model (dashed line without crosses) a mild
SF activity after the QSO phase (i.e. at the beginning of the
passive evolution phase), at a level of 2\% of the peak value and
lasting 1 Gyr. The dot-dashed lines (for clarity these have been
shifted by 0.5 in redshift) refer to objects before the QSO phase
(i.e. in the active star-forming phase), adopting $t_{e}=10$ Myr
(line without crosses), or $t_{e}=50$ Myr (with crosses).}
\label{fig:ero2}
\end{figure}

While the sub-mm counts are informative on the star-formation
phase of the evolution of spheroidal galaxies, the $K$-band counts
(Fig.~\ref{cK}) in the range $14 \lsim {\rm K} \lsim 17$ are
dominated by the passive evolution phase. Fainter than $K \sim 17$
also star-forming spheroids begin to show-up and become
increasingly important at fainter magnitudes. The brighter (in
terms of apparent magnitude) such objects are those of mass lower
than the typical mass probed by present day sub-mm observations,
i.e.\ those having, in the GDS04 model, a prolonged star-formation
phase extending to $z \sim 1$, and whose stars have negligible
obscuration as soon as they come out of their parent MC (see Sect.
\ref{sect:SED}). This is illustrated by Fig.~\ref{cK2}, showing
that, for $K\le 20$, the contribution from star-forming spheroids
peaks at $z\sim 1$. In particular, the predicted number of
spheroids with ongoing star formation at $1.4<z<2$ and $K<20$ is
about twice that of passive spheroids, and their typical stellar
mass is between $5 \times 10^{10} M_\odot$ and $3 \times 10^{11}
M_\odot$, in nice agreement with the findings of Daddi et al.\
(2004; see their Figure 5) and Somerville et al. (2004). The Star
Formation Rate (SFR) of these active objects is $\lesssim 100
M_\odot \mbox{yr}^{-1}$: according to our model, the $K$ band
selects spheroids when their SF (and dust obscuration) is
declining or after it has stopped. Daddi et al.\ (2004) measured a
dust-corrected average SFR of $\simeq 200 M_\odot \mbox{yr}^{-1}$,
higher than our expectations, but still well below the peak SFR of
spheroids according to our scenario. This may suggest that the
present version of our model envisages a too drastic decline of SF
activity in intermediate mass objects (see also Sect.\
\ref{sec:ero}). Note also that Daddi et al.\ adopt a Salpeter IMF
($\propto M^{-1.35}$), while in the GDS04 model we have used a
different IMF ($\propto M^{-1.25}$ above 1 M$_\odot$, $\propto
M^{-0.4}$ below) suggested by chemical constraints (see also
Romano et al.\ 2002). This IMF implies a higher L/M ratio and,
correspondingly, a SFR lower by about 30\%.

We predict that passively evolving spheroids almost disappear at
$K\ge 23$. The observed high-$z$ tails of the distributions for
$K\le 23$ and for $K\le 24$ are thus mostly due to star-forming
spheroids, but the counts are likely dominated by late-type
galaxies, which could account for the big peak of the estimated
(from photometric redshifts) $z$ distribution in the range $0.2
\le z \le 0.9$, as well as for the low-$z$ shoulder at $K \le 20$.

As shown by Fig.~\ref{c6.7std}, according to our model, spheroids
may fully account for the observed $6.7 \mu$m ISOCAM counts and
for the redshift distribution derived by Sato et al. (2004; see
also Flores et al. 1999). Above $\sim 100\,\mu$Jy they are
predicted to be mostly in the passive evolution phase, but the
fraction of active star-forming spheroids increases with
decreasing flux, reaching $\simeq 50$\% at $\simeq 10\,\mu$Jy.

\begin{figure}
\centering
\includegraphics[width=7.5truecm,angle=90]{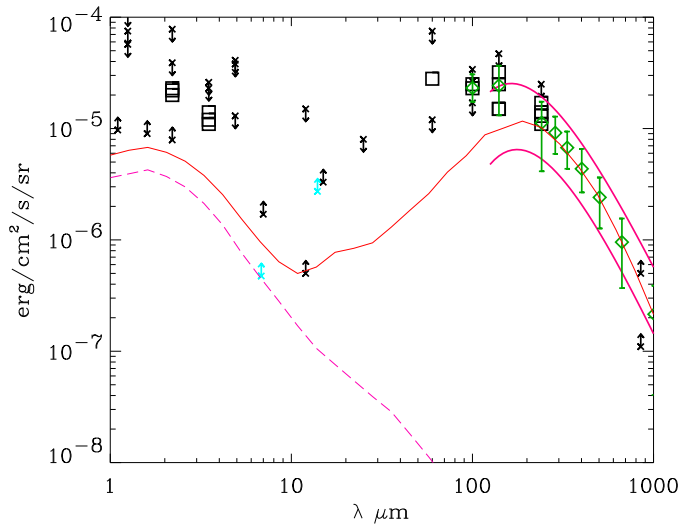}
\caption{Contributions to the $1-1000\mu$m background from
passively evolving plus star-forming spheroidal galaxies (solid
line). Passively evolving spheroids (dashed line) are important at
$\lambda \lsim 15\,\mu$m. Data are from Hauser \& Dwek (2001) and
Metcalfe et al.\ (2003).} \label{bgir}
\end{figure}

\begin{figure}
\centering
\includegraphics[width=9truecm]{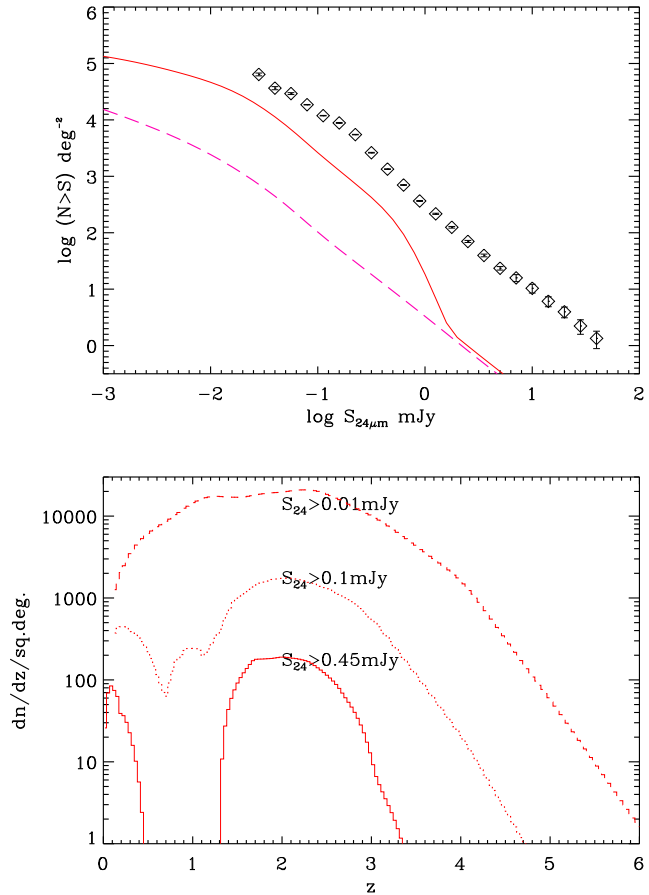}
\caption{{\it Upper panel}: contributions of spheroidal galaxies
to the $24 \mu$m  counts predicted by our reference model compared
with the Spitzer/MIPS data (Papovich et al. 2004). The long-dashed
line shows the counts of passively evolving spheroids. {\it Lower
panel}: predicted redshift distributions for $S_{24\mu{\rm m}}\ge
0.45$ mJy (the SWIRE limit, Lonsdale et al. 2003), 0.1 and 0.01
mJy.} \label{counts_24}
\end{figure}

\begin{figure}
\centering
\includegraphics[width=9truecm]{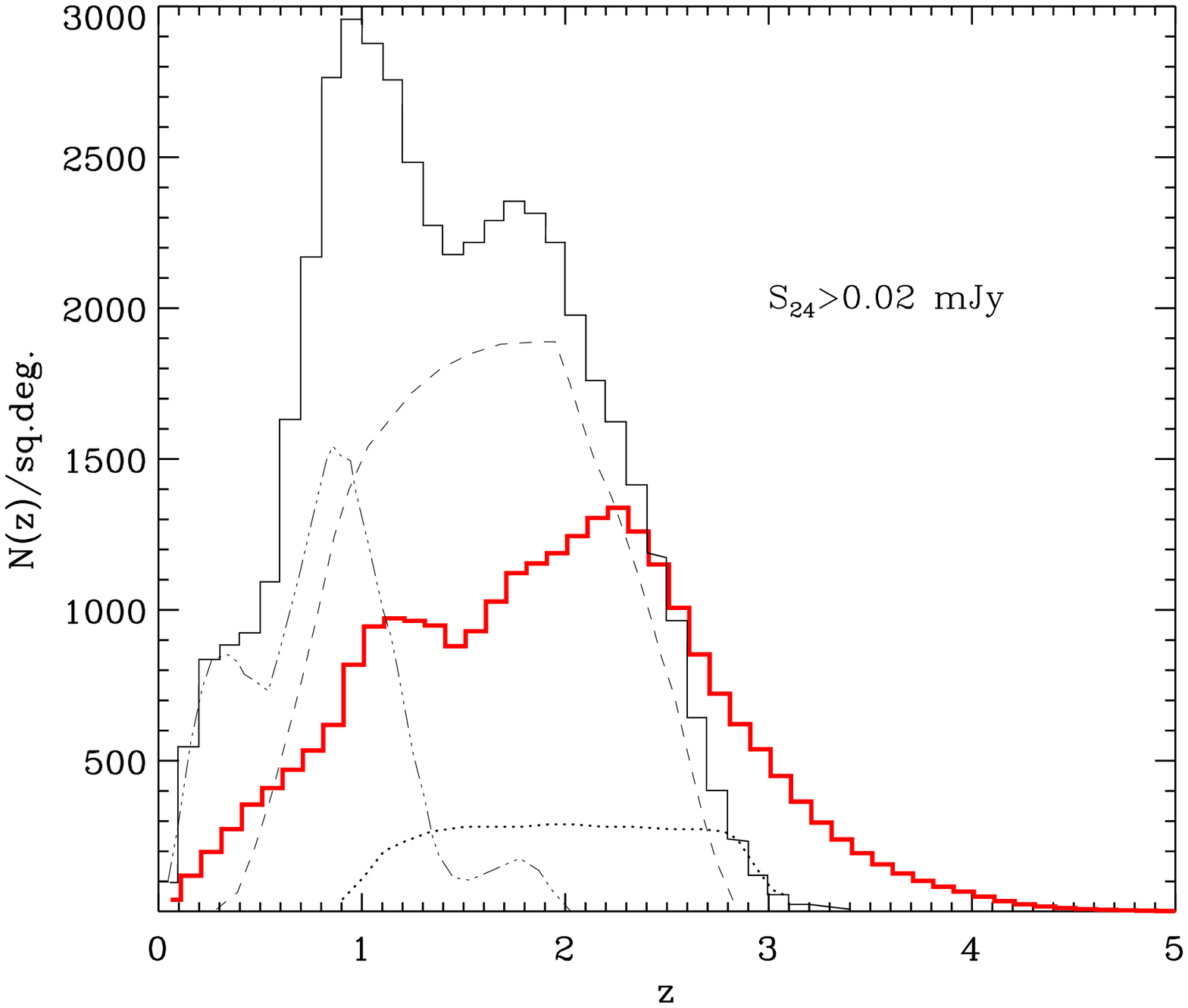}
\caption{Redshift distribution for S$_{24\mu m}>0.02\,$mJy. The
thick histogram corresponds to our reference model for spheroids,
the thin one to the model by Chary et al.\ (2004, their Fig.~4),
which includes the contributions from low luminosity starbursts
(L$_{IR}<10^{11}$ L$_\odot$, 3 dot-dashed), LIRGs with $10^{11}<$
L$_{IR}< 10^{12}$ L$_\odot$ (dashed), and ULIRGs with
L$_{IR}>10^{12}$ L$_\odot$ (dotted).} \label{counts_242}
\end{figure}

\subsection{Extremely Red Objects}
\label{sec:ero}

Extremely red objects (EROs), with $R-K>5$, have received special
attention in recent years (Daddi et al. 2000; Smith et al. 2002;
Roche et al. 2002, 2003; Cimatti et al. 2002, 2003; Takata et al.
2003; Yan \& Thompson 2003; Yan et al. 2004b; Webb et al. 2004;
Moustakas et al. 2004), since their properties set crucial
constraints on the early evolutionary phases of massive spheroidal
galaxies. EROs are actually a mix of dusty star-forming and
evolved galaxies formed at high redshifts (Cimatti et al. 2002,
2003; Mohan et al. 2002).

We have worked out the counts and the redshift distributions of
both star-forming and passively evolving spheroidal galaxies
brighter than $K=19.2$, 20, 20.3 that, according to our model,
have ERO colours (Fig.\ \ref{counts_ero}). This prediction is the
only one, among those presented in this paper, which depends
strongly also on GRASIL parameters other than $\tau_{MC}$ and $k$
(see Sect.\ \ref{sect:SED}). Indeed, the predictions at $z \gtrsim
1$ for dusty and star forming EROs require the computation of rest
frame fluxes at $\lambda \lesssim 3000$ \AA, which are severely
dependent on details of dust obscuration.

In sharp contrast with the standard semi-analytical models, which
severely under-predict this population (e.g., Somerville et al.
2004), our model somewhat over-produces the number of EROs (of
both classes), in the range $1.5 \lsim z \lsim 2$. On the
observational side, the present redshift distributions may be
biased by the strong clustering of these sources and affected by
sampling variance, and some may be incomplete at $z> 1$ (Yan et
al. 2004b). From the theoretical point of view, it should be noted
that a mild star formation activity (say at a level $\sim$ a few
\% of the peak rate), either due to residual gas or induced
occasionally by interactions, neglected by our simple model, could
significantly decrease the predicted number of passive galaxies
with ERO colours, particularly at high redshifts. To illustrate
this possibility, Fig.\ \ref{fig:ero2} shows the effect on
predicted z-distribution of EROs brighter than $K=20$ of an
'artificial' (i.e.\ not predicted by the model equations) episode
of star formation starting after the feedback from the QSO has
swept out the ISM (QSO phase in the terminology of GDS04), and
lasting 1 Gyr. A decline of the SFR somewhat less abrupt than
implied by the GDS04 model seems also to be suggested by the
results of Daddi et al.\ (2004), who find that star forming
objects at $1.4<z<2$ in the K20 survey have a dust-corrected star
formation somewhat higher than our prediction (Sect.\
\ref{sect:coured}). As for dusty and star forming EROs, we
reminded above that our prediction depends on details of dust
obscuration. For instance a shorter $t_{e}$ and/or larger values
of the galaxy scale length $r_c$ would translate in less reddened
objects, without affecting significantly the other results
presented in this paper. An example of this is also shown in Fig.\
\ref{fig:ero2}.

In conclusion, it is encouraging that our model has not troubles
in producing {\it enough} high-z objects with ERO colours. More
statistic in the observations will allow us to improve the
treatment of some aspects of the model, such as the amount of star
formation after the QSO phase and the dust obscuration.

\begin{figure}
\centering
\includegraphics[width=8truecm]{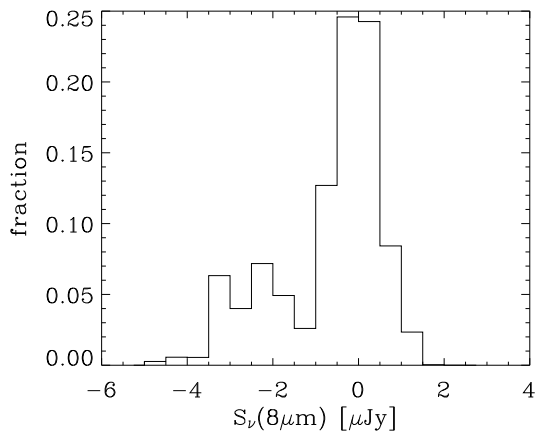}
\caption{Expected distribution of the $8\,\mu$m flux density of
our model spheroids brighter than $100\,\mu$Jy at $24\,\mu$m
(reference model).} \label{frac8}
\end{figure}

\subsection{The IR background}
\label{sec:irb}

A further constraint on the model comes from the IR background. In
Fig.~\ref{bgir} we compare the $1-1000\mu$m background spectrum
yielded by our model with the data collected by Hauser \& Dwek
(2001). Longward of $\lambda \simeq 200\,\mu$m star-forming
spheroids alone may explain the observed background, while at
shorter wavelengths, and, in particular, at the intensity peak,
late-type (especially starburst) galaxies are expected to
dominate.

\subsection{Contribution to $24 \mu$m Spitzer counts}
 \label{sec:c24}

Figure~\ref{counts_24} shows the contribution from forming
spheroids to the $24\,\mu$m number counts, and the predicted
redshift distributions at various flux limits.

In particular, our reference model predicts a non-negligible
contribution, $\simeq 20$\%, to the $24\,\mu$m Spitzer counts
above $\simeq 0.11\,$mJy. Yan et al.\ (2004a) report that 13\% of
the detected sources do not have a counterpart at both $8\,\mu$m
and $0.7\,\mu$m, to flux limits of 20 and $0.17\,\mu$Jy
respectively. In our view, this sub-population should be dominated
by forming spheroids. Figure~\ref{frac8} reports the distribution
of $S_\nu(8\mu\mbox{m})$ for our model population with
$S_{24\mu{\rm m}} \gtrsim 100 \, \mu$Jy, and shows that only few
of these sources have $S_{8\mu{\rm m}}> 20\,\mu$Jy. We expect that
the same is true at $0.7\,\mu$m, although predictions at this
short wavelength, probing the rest frame UV for our model sources,
are very uncertain.

In Fig.\ \ref{counts_242}, we have also compared the redshift
distribution predicted by Chary e al.\ (2004, their Fig.~4) for
$S_{24\mu{\rm m}}>0.02\,$mJy with the one from our model for
spheroids. Chary et al. take into account low luminosity
starbursts, LIRGs and ULIRGs. Note that our model leaves room for
(but does not include) populations other than forming spheroids
only at redshift $\lesssim 1.5$; therefore we predict much fewer
sources at high redshift as compared to Chary et al., but with a
tail extending to higher z. Also according to our scenario, the
populations responsible for the bulk of the 15 and $24\,\mu$m
counts should be at redshift $\lesssim 1.5$, since all massive
halos virialized at higher z are used to build spheroids (see
GDS04 for more details). The observed $24\,\mu$m redshift
distribution will be a strong test for our model.

\subsection{Contribution to Spitzer-IRAC counts}
 \label{sec:cirac}

\begin{figure*}
\centering
\includegraphics[width=16truecm]{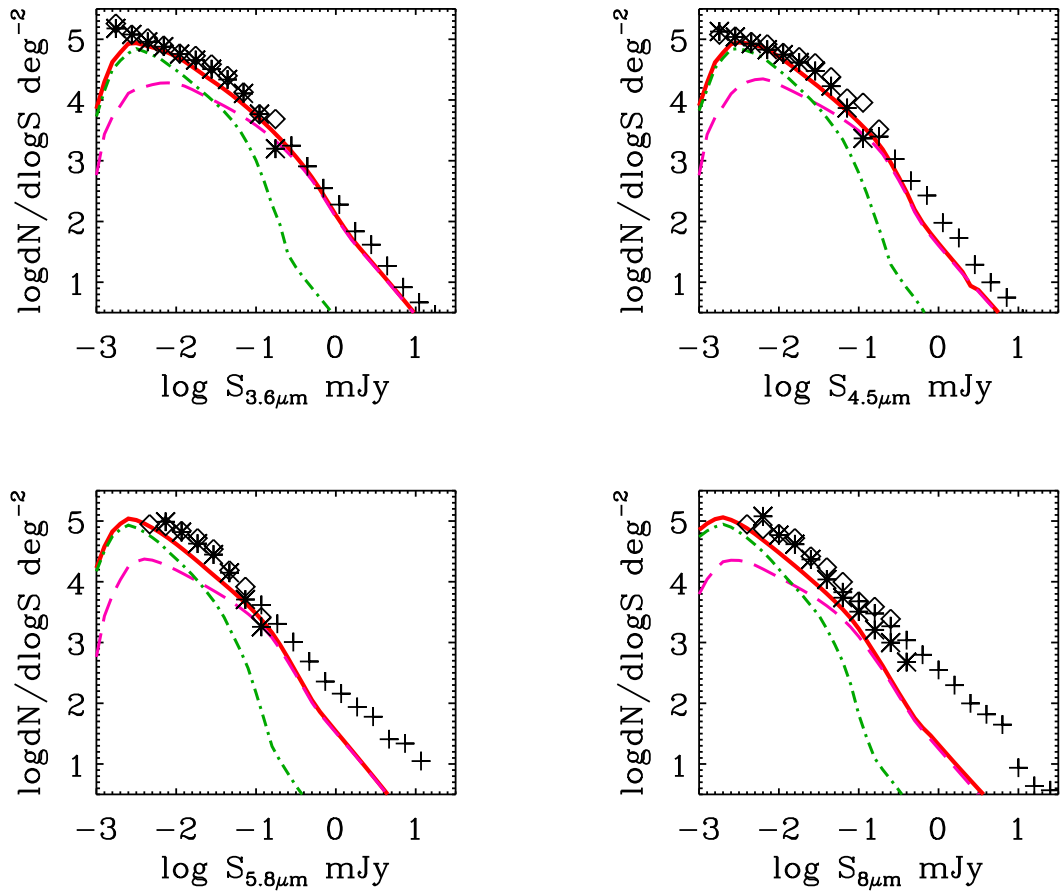}
\caption{Contributions of spheroidal galaxies to the counts in the
Spitzer/IRAC bands predicted by our reference model shown in
comparison with data from Fazio et al.\ (2004). The thick
continuous line is the total, the long dashed and dot-dashed lines
are for passive and star-forming spheroids, respectively. }
\label{countsirac}
\end{figure*}

In Fig.\ \ref{countsirac} we show the contribution of spheroidal
galaxies to the Spitzer/IRAC bands, as predicted by our reference
model. As expected based on results for the $K$ and the $6.7\mu$m
bands, spheroids dominate the counts (except perhaps at the bright
end), with a contribution decreasing from the near- to the mid-IR.
Spheroids in the active star-forming phase dominate at faint flux
density levels ($\lesssim 30 \mu$Jy).

%%%%%%%%%%%%%%%%%%%%%%%%%%%%%%%%%%%%%%%%%%%%%%%

\section{Discussion and conclusions}
\label{sect:conclu}

Granato et al. (2001, 2004a) have shown that the mutual feedback
between star-forming spheroidal galaxies and the active nuclei
growing in their cores can be a key ingredient towards overcoming
one of the main challenges facing the hierarchical clustering
scenario for galaxy formation, i.e. the fact that the densities of
massive high redshift galaxies detected by SCUBA and by deep
near-IR surveys are well above the predictions.

The model by GDS04 interprets the SCUBA galaxies as
proto-spheroids in the process of forming most of their stars. In
fact, according to this model, essentially all massive halos
virializing at $z\gsim 1.5$ give rise to elliptical galaxies or to
bulges of later type galaxies, that form at lower redshifts. The
star-formation in small halos is suppressed by supernova-driven
winds, so that the early star-formation occurs primarily in the
more massive halos, through gigantic starbursts with star
formation rates $\sim 10^3\,M_\odot/$yr, whose duration is
determined by the feedback from the active nuclei, followed by
essentially passive evolution. The model can therefore be
appropriately tested primarily against surveys selecting high-$z$
galaxies, like the (sub)-mm surveys by SCUBA and MAMBO, or
spheroidal galaxies, like near to mid-IR surveys.

To this end we first assessed the values of the adjustable
parameter $\tau_{\rm MC}$ of the spectrophotometric code GRASIL
(Silva et al.\ 1998), which regulates the effect of the complex
and poorly understood radiative transfer processes on the
time-dependent SEDs. To properly deal with the extreme conditions
prevailing during the intense star-formation phase of spheroidal
galaxies, a new parameter, $k$, had to be introduced. Having fixed
the values of the two parameters,  and keeping the GDS04 choice
for the parameters controlling the star-formation history, the
chemical enrichment and the evolution of the dust and gas content
of massive spheroidal galaxies, we have shown that the model
accounts, on one side, for the SCUBA and MAMBO counts,
corresponding to the active star-forming phase, and, on the other
side, for the $K$-band counts for $K \lsim 20$--21 and for the
redshift distribution of $K < 20$ galaxies, as well as for the
high-$z$ tails of redshift distributions up to $K=24$.

Although observations unambiguously indicate that stellar
populations in nearby spheroidal galaxies are old, it is possible
that stars were formed in small sub-units, before the galaxy was
assembled. Our results highlight a clear continuity between
objects where stars formed (detected by (sub)-mm surveys) and the
evolved galaxies dominating the bright $K$-band counts, indicating
that massive spheroidal galaxies formed most of their stars when
they were already assembled as single objects. As illustrated by
Fig.~\ref{cK2}, the model is remarkably successful in reproducing
the observed $K<20$ redshift distribution for $z>1$, in contrast
with both the classical monolithic models (which overestimate the
density at high-$z$) and the semi-analytic models (that are
systematically low). For example, Somerville et al. (2004) find
that their hierarchical model underproduces $K\le 20.15$ ($K_{AB}
\le 22$ in their notation) galaxies by about a factor of 3 at
$z\gsim 1.7$, and by an order of magnitude at $z\gsim 2$, while a
monolithic model overproduces these galaxies by about a factor of
2 at $z\sim 2$. The GDS04 model also reproduces the bimodal
distribution of colours of the Somerville et al. (2004) sample, as
due to contributions of star-forming and passively evolving
spheroids. The ratio of star-forming to passively evolving
spheroids is also nicely consistent with the finding of Somerville
et al. (2004) and Daddi et al. (2004), although the observations
of the latter authors suggest a more gradual decline of the
star-formation rate than implied by the current version of our
model. Again, reproducing the observed colour distribution proves
to be very challenging for both monolithic and standard
semi-analytic models.

A specific analysis of the $K$-band counts and redshift
distributions of EROs has been worked out, since these objects
appear to provide particularly sharp constraints on galaxy
evolution models. The results confirm that the GDS04 model
successfully reproduces the counts, and is in reasonably good
agreement with the redshift distributions, although the need of a
more sophisticated modelling of the star formation history and of
dust geometry is indicated.

The mid-IR counts provided by ISOCAM, primarily at 6.7 and
$15\,\mu$m, and by the Spitzer Space Telescope at several
additional wavelengths and over larger areas, constrain the SEDs
of spheroidal galaxies which dominate for $\lambda \lsim 6\,\mu$m,
and the evolution properties (as well as the SEDs) of late-type
(normal and starburst) galaxies that dominate at longer
wavelengths. The ISOCAM counts at 6.7$\mu$m and the associated
redshift distribution are well fitted by the model
(Fig.~\ref{c6.7std}), as are the Spitzer/IRAC counts at 3.6, 4.5,
and $5.8\,\mu$m.

At longer wavelengths, we expect that the contribution of
starburst galaxies to the counts becomes increasingly important,
particularly at bright flux densities, and to actually dominate at
$24\,\mu$m (Fig.~\ref{counts_24}). Although modelling the counts
of such sources is beyond the scope of the present paper, we note
that, according to the GDS04 model, essentially all {\it massive}
halos at $z\gsim 1.5$ turn into spheroidal galaxies. As a
consequence, we expect a sharp drop in the redshift distribution
of starburst galaxies at $z\gsim 1.5$, in contrast with
predictions of some current phenomenological models (Dole et al.
2003; Chary \& Elbaz 2001), and a high-$z$ tail due to
star-forming spheroids. For example, we expect a surface density
of sources with $S_{24\mu{\rm m}} > 20\,\mu$Jy and $z\simeq 2$
about 2 times lower than predicted by Chary et al. (2004, their
Fig.~4).

While the scanty redshift information on $24\,\mu$m sources (Le
Floc'h et al. 2004) does not allow us to test our expectations
yet, such sources can be preliminarily characterized by their IR
colours (Yan et al. 2004a). Remarkably, the predicted fraction of
star-forming spheroids with $8\,\mu$m fluxes below the detection
limit matches that found by Yan et al. (2004a). We thus interpret
these sources as proto-spheroidal galaxies in the process of
forming most of their stars, at typical redshifts $\simeq 2$ but
with a tail extending up to $z\simeq 4$.

Moving to still longer wavelengths, we expect that star-forming
spheroids start dominating the $160\,\mu$m counts below
50--$100\,$mJy, and produce a peak at $z\sim 1.5$ in the redshift
distribution, while at brighter fluxes, counts are likely
dominated by low redshifts sources (Rowan-Robinson et al. 2004).

It is worth noticing that the AGN activity powered by the growing
SMBHs has a small effect on the number counts of {\it spheroids}
at all wavelengths considered here. Since the AGN SED ($\nu
L_\nu$) is relatively flat at optical to far-IR wavelengths and
drops in the sub-mm region, the maximum AGN contribution occurs
around $15\,\mu$m, at the minimum between the peak of direct
starlight emission and that of dust emission (Silva, Maiolino \&
Granato 2004; see also Fig.~\ref{bgir}). But, at this wavelength,
the spheroids are anyway a minor component of the counts (see the
upper left-hand panel of Fig.~\ref{c15_850}). The sub-mm counts
are dominated by proto-spheroids with very high SFRs ($\hbox{SFR}
\gtrsim 500\, M_\odot\,$yr for sources brighter than 5 mJy), while
the accretion rate onto the SMBH predicted by the model is several
orders of magnitude lower (see Figs. 2 and 3 of GDS04). Therefore
the AGN luminosity is generally much lower than that due to stars
(it is non-negligible only for a very small fraction of the star
forming phase) even allowing for a much higher mass to light
conversion efficiency.  On the other hand, we expect the mild AGN
activity to be detectable in X-rays in the majority of the bright
SCUBA sources (Granato et al. 2004b), in keeping with the findings
by Alexander et al. (2003). A detailed analysis will be presented
in a forthcoming paper (Granato et al.\ in preparation). The
counts of spheroids at $\lambda <10\,\mu$m are dominated by
objects in the passive evolution phase, for which the mass
accretion rates onto the SMBHs are very low and, correspondingly,
the AGN activity is very weak.

\section*{Acknowledgments}

Work supported in part by MIUR (through a COFIN grant) and ASI. LS
and GLG acknowledge kind hospitality by INAOE where part of this
work was performed, as well as support by the LSGLG foundation.
Finally, we wish to thank an anonymous referee for helping us in
improving the presentation of this work.

 \label{lastpage}

\end{document}